\documentclass[preprint,proceedings]{rmaa}
\newcommand{\Ha}{H$\alpha$}
\newcommand{\Hb}{H$\beta$}
\newcommand{\htwo}{H$_{2}$}

\newcommand{\kms}{km~s{$^{-1}$}}

\newcommand{\HaNII}{H$\alpha$ + [N~II]}

\usepackage{paralist}

\usepackage{psfrag,color}

\SetYear{2005}
\SetConfTitle{The Ninth Texas-Mexico Conference on Astrophysics}

\title{The 3-D Structure of the Helix Nebula}

\author{
  C. R. O'Dell\altaffilmark{}}

\altaffiltext{1}{Vanderbilt University, Nashville, TN.}

\shortauthor{O'Dell}
\shorttitle{3-D Structure of the Helix Nebula}

\fulladdresses{
\item C. R. O'Dell: Department of Physics and Astronomy, Box 1807-B, Vanderbilt University, Nashville, TN 37235 (\email{cr.odell@vanderbilt.edu}).
}

\listofauthors{C. R. O'Dell}
\indexauthor{C. R. O'Dell}

\abstract{The 3-D structure of the Helix Nebula has been addressed multiple times and is slowly yielding to
application of monochromatic emission line imaging, high resolution spectroscopy, and photoionization theory.
The inner structure of the nebula is a toroidal disk filled with doubly ionized helium and is ionization bounded, with an
extended vertical component resembling the structure of numerous bipolar nebulae. The outer structure is less well defined,
with one construction being that there is an open-center outer-disk that is nearly perpendicular to the inner-disk
and an alternate construction that explains the features seen just outside the inner-disk as being the 
result of the observer's line of sight passing through extended lobes perpendicular to the inner-disk. A definitive
model awaits thorough velocity mapping in the major diagnostic emission lines. However, even our well-defined knowledge of the inner-disk defies explanation by the simplest application of the broadly accepted two-wind model for the formation of PN.}

\addkeyword{ISM: Planetary Nebulae--Planetary Nebulae: individual (NGC 7293)}

\begin{document}
\maketitle

\section{Introduction}
The problem of mapping objects seen in two dimensions (in the plane of the sky) into three dimension (3-D) confronts the 
astronomer often but in no case is this more important than in the study of the shape of the Planetary Nebulae (PN). If we 
understand their 3-D form we can hope to decipher the method of their formation and project their future involvement in the
interstellar medium (ISM). One can argue that forms with circular symmetry have a rotational axis and the radial velocity
distribution within a feature can lead to understanding its 3-D form. When we combine this with the fact that the PN are
dominated by photoionization, this means that we have an additional tool in 3-D modeling, for we know the invariable progression
with distance of different ionization states which give rise to the emission lines forming the images. It is a rare case 
when one has the combination of detailed imaging in all of the major diagnostic lines and complete radial velocity maps.
The Helix Nebula is no exception to this rule-of-incompleteness of data, but, because it is the nearest PN it merits 
investigation using the information that one can derive. That information comes slowly because it is one of the lowest
surface brightness classic PN.

The basic structure of the innermost portion of the Helix Nebula was independently demonstrated in 1998 to be a disk inclined slightly
to the plane of the sky (Meaburn et~al. 1998, henceforth M98; O'Dell 1998, henceforth O98), although only the latter paper demonstrated that the inner
part of the disk appears ``empty'' only because it is a zone of doubly ionized helium which appears only in HeII emission.
O'Dell, McCullough, \&\ Meixner (2004, henceforth OMM04) presented a new level of imaging in multiple emission lines and derived a 3-D model composed of a refined view of the inner-disk and argued that the outer part of the nebula could be explained by 
there being a surrounding flat disk of material lying almost perpendicular to the inner-disk. They argue that the inner disk has a perpendicular low-density extension that lies along its symmetry axis and forming the plumes that lie to the northwest and southeast of the nebula.  Meaburn \etal\ (2005, henceforth M05) revisited 
the question of the 3-D structure and 
\begin{figure*}
\includegraphics[width=\textwidth]{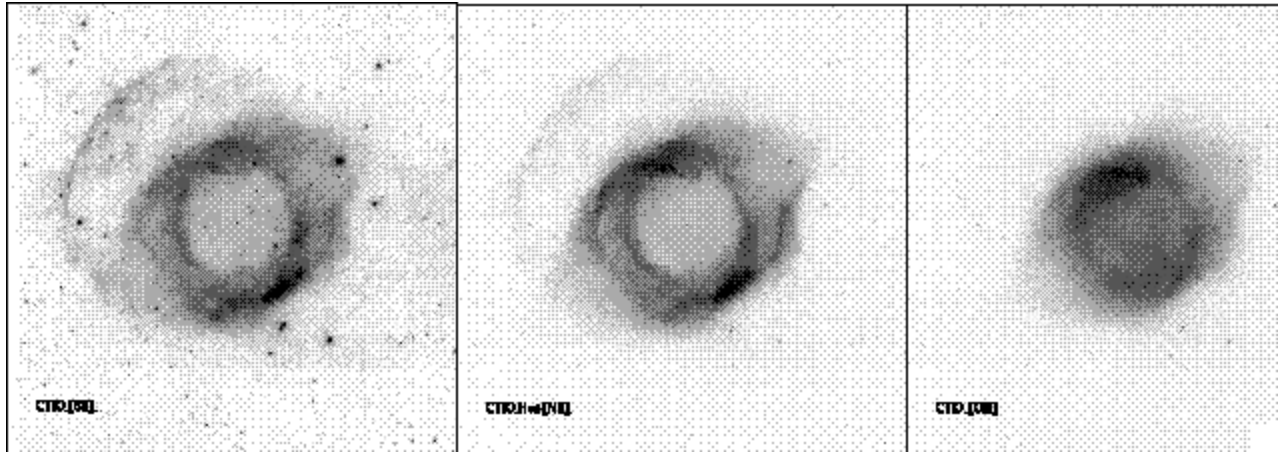}
\caption{These three images of the Helix (OMM04) show the dramatic difference of the appearance of the nebula
in different stages of ionization. The field of view is 1761\arcsec x1761\arcsec\ in each.}
\label{fig:simple1}
\end{figure*}
argue that the inner part of OMM04's outer-disk is actually caused by lobes that
are perpendicular to the inner-disk, the outer-ring that is seen on images being caused by where the line-of-sight passes
through significant amounts of material in these lobes.  For reference in this article we show a portion of the OMM04 emission line images in Figure 1 and a sketch in Figure 2 that highlights the major features.

One knows that the inner-disk is optically thick to Lyman continuum (LyC) radiation because it shows the 
expected progression of ionization states from HeII through [S~II].  The feature labeled ``outer-ring'' in Figure 2 shows
a similar progression from [O~III], [N~II], through [S~II], indicating that it too is ionization bounded. The ionization of the outer-disk (from the outer-ring to the outermost-border) would require that material to be ionized by scattered LyC photons from the inner-disk or it is fossil ionization, i.e. low density material photoionized earlier and still recombining.  The magnitude of a 
a surrounding molecular cloud is uncertain as one does not see it in the highest resolution molecular images (in \htwo, 
Speck \etal\ 2002) and the radio studies of CO (Young \etal\ 1999) and HI (Rodr\'{\i}guez, Goss, \&\ Williams 2002) are of low spatial resolution. However, their high velocity resolution
indicates that the molecular radiation comes from small volumes, which are probably the knots which populate the inner-disk
and the outer-ring (OMM04).  
\section{Available Information}
Fortunately, the Helix has been well imaged by this time. Deep monochromatic emission line images of the entire nebula at about 1\arcsec\ 
resolution exist (OMM04) in \HaNII, [S~II], [O~III], and \Hb, in addition to 2.3\arcsec\ images in HeII (O98), and a very deep image in \HaNII\ at 4.1\arcsec\ (M05). \htwo\ images at 2\arcsec\ (Speck \etal\ 2002) cover the brightest part of the
nebula and radio surveys in CO (Young \etal, 1999) at 31\arcsec\ exist, in addition to HI at about 54.2\arcsec x39.3\arcsec\  (Rodr\'{\i}quez
 \etal\ 2002). High resolution spectroscopy has covered only parts of the nebula in the optical lines (M98, O'Dell \etal\ 2002, M05, and of course the radio images also provide high velocity resolution spectra, although at low spatial resolution.

These images are largely complementary and confirming of one another, with one exception. M05 indicate that their images
at arcsecond resolution (in \HaNII) show radial ``spokes'' close to the central star.  However, they caution that those images were not flat-field corrected. Also, they were made with a single CCD, factors making it difficult to reliably see low contranst features. The high S/N images of OMM04, which were made with multiple CCD's that 
were individually flat-field corrected, do not show these features.
\section{Structure of the Main Disk of the Nebula}
The inner-disk appears as an ellipse of 499\arcsec x459\arcsec\ major axes (OMM04). 
A compilation of all available radial velocities allowed OMM04 to determine that it is a circular disk
with the northwest portion closer to the observer and that the plane of the disk is inclined 28\arcdeg\ from the sky (M05 give slightly different
\begin{figure}[!t]
\includegraphics[width=\columnwidth]{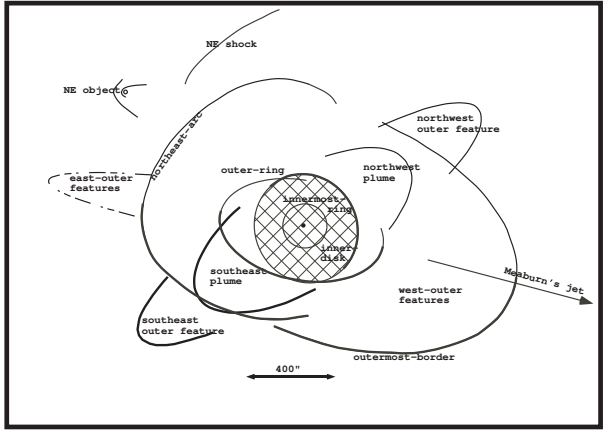}
\caption{This sketch shows the location of the main features of the Helix Nebula discussed in the text. The near side features are shown in heavier lines, e.g. the northwest portion of the inner-disk.}
\label{fig:simple2}
\end{figure}
numbers). OMM04 derive a spatial expansion velocity of 40 \kms\ and from this a formal age of 
6,560 years. An unexplained feature within the inner-disk is something labeled ``innermost-ring'' in Figure 2. This feature is visible only in [O~III], peaks at a diameter of 200\arcsec, and is an enhancement of about 20\%\ in the surface brightness.
It occurs near the outer limits of the HeII core, which is nearly Gaussian in form with a full width at half maximum of 277\arcsec. It was originally identified in M98.
\section{Structure Perpendicular to the Main Disk}
There certainly is structure perpendicular to the main or inner-disk. The inner parts of these are seen as ``plumes''
extending to the northwest and southeast. There are outer features (OMM05) not shown in the depiction in Figure 1 (but are drawn in Figure 2) which
are aligned with the plumes and their parabolic shape indicate that they are either shocks or an extension of the bipolar 
forms seen in numerous smaller PN. After consideration of the tilt angle, the outer parts extend to 3.9 times the inner-disk's diameter above and below the plane of the inner-disk.

\section{Alternative Models for the Outer Ring}
The reader will have noted that I've used the term inner-disk and main disk interchangeably. This is because in OMM04 it was 
argued that outside of the inner-disk there is an outer-disk, composed of an ionization bounded outer-ring surrounded by
a thin disk of low density material extending out to the outermost-border in Figure 2. In this model the arcuate features
outside of the inner-ring are interpreted as an incomplete ring of material tilted 53\arcdeg\ out of the plane of the sky
and oriented with the southeast side being closest to the observer. This structure is suggested by the fact that there
clearly is an ionization boundary along those arcuate structures, they are marked by young knots arising from an ionization
front, and their velocity distribution with position angle (PA) resembles that of an expanding ring. The summary of optical and
radio velocities of the nebula and the knot given in Figure 15 of OMM04 shows evidence for the sinusoidal variation of
radial velocity that is expected, although the scatter of the velocities is large. These features were first identified in the CO (Young \etal\ 1998)studies, where two kinematic systems were first seen, one being smaller (now associated with the inner-disk) and the
other further out. In this interpretation the outer-disk is nearly perpendicular to the inner-disk, with the vertical structure associated with the inner-disk interrupting portions of the outer-disk's innermost feature, the outer-ring. This
outer-ring portion of the outer-disk would have a diameter of 742\arcsec, an expansion velocity of 32 \kms, and a dynamic age of 12,100 years.

A quite different interpretation of the outer-ring features appears in M05, where they argue that these are composed of views
through lobes extending above and below a line perpendicular to the inner-disk. This multi-free-parameter model succeeds
in producing an equally good explanation of the appearance of the nebula and predicts velocity patterns similar to what
is observed. East-west and north-south p-v diagrams would be similar for both models. I believe that the one discriminator
between the two models is that in the interval PA=80\arcdeg -150\arcdeg, the ``lobes'' model would predict the radial velocity of the near-side component
would be nearly constant with PA, while the data in OMM04 (Figure 15) show that the radial velocity becomes rapidly 
more negative in this interval.

\section{The Outermost Features Associated with the Nebula}
There are numerous features that fall outside the outer-ring in addition to those I identify with material that is 
perpendicular to the inner-disk. The largest is the material I associate with the outer-disk, which is a series of 
irregular structures bounded by the outermost-border in Figure 2.  The general similarity of form suggests, but does not 
establish an association with the outer-ring feature, this similarity being the only argument for an outer-disk, rather than there
simply being an outer-ring.  In any event, the northeast-arc feature is almost certainly a shock caused by the nebula
moving through the ISM (OMM04).  The features labeled ``east-outer features'' in Figure 2 were first reported by Kwitter
\etal\ (1993)
and it was argued by Borkowski (1993) that these were shocks created by interaction of the outer nebula with the ISM. If there is a collimated
flow of material in that direction, then those may be a source of the shocks. There are additional shocks (NE object and NE
 shock) to the northeast, but their origin is uncertain as these shocks have no symmetry along a radius vector, although
there is an approximate symmetry axis lying close to the direction of the nebula's proper motion (OMM05).  

In their low spatial resolution, but star-subtracted \HaNII\ image 
M05 report the discovery of a ``jet'' beginning within the outermost-border and extending beyond towards PA=255\arcdeg.
M05 say that there may be a similar but opposite feature just outside of the outer-ring and that this may be the driving
material for the shocks in the ``east-outer features''. Meaburn's jet is easily visible in the broadband (1344-2831 \AA) GALEX images (http://www.galex.caltech.edu/MEDIA/2005-02), but nothing well identified with a counter-jet is seen. The ``east-outer features'' lie at PA=81\arcdeg, which is not 
opposite to Meaburn's jet, which reduces the likelihood of an association, leaving both these shocks and Meaburn's jet for
explanation in the future.
\section{Stratified Flow in the Direction of the Center of the Nebula}
Even without a detailed kinematic model for the entire nebula, we can identify some important velocity features. A sample 
across the central 100\arcsec\ diameter shows several interesting features. O'Dell \etal\ (2002) found central values of the line splitting of 32 \kms\ in \Ha, 49 \kms\ in [N~II], and M05 find 25 \kms\ for [O~III]. M05 argue from observations 
of the intrinsically faint HeII 6560 \AA\ line that the HeII line splitting is less than 24 \kms\ while my study (in progress) of the intrisically much stronger HeII 4686 \AA\ line 
shows a well defined splitting of 18 \kms. The progression of increasing splitting velocity with decreasing ionization 
state (c. f. the review in O98 for details) is well defined HeII 18 \kms, [O~III] 25 \kms, [N~II] 49 \kms\ (\Ha\ comes from
each ionization zone but is weighted towards the cooler, outer parts of the [O~III] zone and has the splitting of 32 \kms).
This is the same pattern recognized more than fifty years ago (Wilson 1950). This ``Hubble'' type flow runs contrary to the expectations of the widely adopted two-wind model for the origin of the forms of the PN
as does the presence of a filled central region that we see in both the Helix and the Ring Nebulae (O'Dell \etal\ 2002). 
Perhaps the two-wind model applies in the youngest PN, but in the older objects we are seeing other effects, such as filling-in following the cessation of the driving high velocity stellar wind.
However, one should not over-interpret the above velocity trends because they are determined from looking down onto the
inner-disk of the nebula at an angle of about 23\arcdeg\ (OMM04) and we know that it has significant vertical structure.
What is needed to test the two-wind hypothesis are detailed velocity-based models of PN over a variety of ages and orientations.
In fact, it is this kind of detailed observation that is necessary for removing the ambiguity of the interpretation of the
outer-ring of the Helix nebula and is the subject of a continuing investigation by the author.

\acknowledgements
I am grateful to Will Henney and John Meaburn for reading and commenting on a draft of this paper, although conflicting 
viewpoints with John still remain.
This work was supported in part by a Space Telescope Science Institute 
grant supporting program GO-9944.

\end{document}